\documentclass[10pt,preprint]{aastex}

\newcommand{\myemail}{keivan.stassun@vanderbilt.edu}
\newcommand{\bd}{2M0535--05}
\newcommand{\model}{{\sc wd}}

\shorttitle{Brown-Dwarf Eclipsing Binary}
\shortauthors{Stassun et al.}

\begin{document}

\title{A Surprising Reversal of Temperatures in the Brown-Dwarf
Eclipsing Binary 2MASS J05352184$-$0546085}

\author{Keivan G.\ Stassun\altaffilmark{1},
Robert D.\ Mathieu\altaffilmark{2}, and
Jeff A.\ Valenti\altaffilmark{3} }

\altaffiltext{1}{Department of Physics \& Astronomy,
Vanderbilt University, Nashville, TN 37235; \myemail}
\altaffiltext{2}{Department of Astronomy, University of
Wisconsin--Madison, Madison, WI, 53706} 
\altaffiltext{3}{Space Telescope Science Institute, Baltimore,
MD 21218}

\begin{abstract}
The newly discovered brown-dwarf eclipsing binary 
\object{2MASS J05352184$-$0546085} provides a
unique laboratory for testing the predictions of theoretical models of
brown-dwarf formation and evolution. The finding that the lower-mass
brown dwarf in this system is hotter than its higher-mass companion
represents a challenge to brown-dwarf evolutionary models, none of
which predict this behavior. Here we present updated determinations
of the basic physical properties of \bd, bolstering the surprising
reversal of temperatures with mass in this system. We compare these
measurements with widely used brown-dwarf evolutionary tracks, and
find that the temperature reversal can be explained by some models
if the components of \bd\ are mildly non-coeval, possibly consistent
with dynamical simulations of brown-dwarf formation. Alternatively,
a strong magnetic field on the higher-mass brown dwarf might explain
its anomalously low surface temperature, consistent with emerging
evidence that convection is suppressed in magnetically active, low-mass
stars. Finally, we discuss future observational and theoretical work
needed to further characterize and understand this benchmark system.
\end{abstract}


\section{Introduction\label{intro}}
Born with masses between the least massive stars 
\citep[$\approx 0.072\; {\rm M}_\odot$;][]{chabrier00}
and the most massive planets 
\citep[$\approx 0.013\; {\rm M}_\odot$;][]{burrows01}, 
brown dwarfs at once extend our
understanding of the formation and evolution of both stars and planets.
In the decade since the discovery of the first brown dwarfs
\citep{nakajima95,rebolo95},
the study of star- and planet-formation has increasingly trained its
attention on these objects that, while being neither star
nor planet, may provide key insights to understanding the origins
of both \citep{basri00}.

Such an understanding must be founded on accurate measurements
of fundamental physical properties---masses, radii, and
luminosities. Unfortunately, the number of objects for which these
properties can be measured directly is extremely small. Though
rare, eclipsing binaries have long been employed as ideal
laboratories for directly measuring fundamental stellar parameters
\citep[e.g.][]{andersen91}. The power of eclipsing binaries lies
in their provision of masses and radii with only the most basic of
theoretical assumptions. With the addition of atmosphere models the
ratio of effective temperatures can also be derived, and luminosities
can then be determined directly through the Stefan-Boltzmann law
without knowledge of distance. Finally, considering the two objects'
properties together permits study of the binary as twins at birth
whose evolutionary histories differ because of their different masses.

Thus the recent discovery of a brown-dwarf eclipsing binary system
in the Orion Nebula Cluster, \object{2MASS J05352184$-$0546085}
(hereafter \bd), offers a unique laboratory with which to
directly and accurately test the predictions of theoretical
models of brown-dwarf formation and evolution \citep[][hereafter
Paper~I]{stass06}. In Paper~I, we presented a preliminary analysis of
the orbit and of the mutual eclipses in \bd\ to directly measure the
masses and radii of both components, as well as the ratio of their
effective temperatures. Our mass measurements reveal both objects in
this young binary system to be substellar, with masses of $M_1 = 55$
M$_{\rm Jup}$ and $M_2 = 35$ M$_{\rm Jup}$, accurate to $\sim 10\%$. In
addition, from the observed eclipse durations and orbital velocities
we directly measured the radii of the brown dwarfs to be $R_1 = 0.67$
R$_\odot$ and $R_2 = 0.51$ R$_\odot$, accurate to $\sim 5\%$, and
representing the first direct measurements of brown-dwarf radii. Such
large radii are generally consistent with theoretical predictions of
young brown dwarfs in the earliest stages of gravitational contraction.

Surprisingly, however, we reported in Paper~I that the lower-mass
brown dwarf has an effective temperature that is slightly---but
significantly---warmer than its higher-mass companion.  Such a reversal
of temperatures with mass is not predicted by any theoretical model
for coeval brown dwarfs. This finding has potentially important
ramifications for theoretical brown-dwarf evolutionary tracks, and
thus for our understanding of brown-dwarf formation more generally.

In this paper, we present supplementary light-curve and radial-velocity
measurements (\S\ref{data}) which we use to refine our determination
of the basic physical properties of \bd\ (\S\ref{results}). We verify
the finding of a temperature reversal with mass in this system, and
quantitatively compare the empirically determined physical parameters
with those predicted by theoretical evolutionary tracks. We then
briefly explore the implications of the temperature reversal in
\bd\ for theoretical models of brown-dwarf formation and evolution
(\S\ref{discussion}). Specifically, we consider whether the temperature
reversal can be explained by dynamical brown-dwarf formation scenarios
and/or by the physical effects of strong surface fields on young
brown dwarfs. We summarize our conclusions in \S\ref{conclusions}.

\section{Data and Methods\label{data}}

\subsection{Spectroscopic observations: Radial velocities
and spectral types\label{spectra}}
We observed \bd\ with the {\it Phoenix} spectrograph on Gemini South
on eight separate nights from December 2002 to January 2003 (first
reported in Paper~I) and again on 03 March 2005 (newly reported
here). All nine observations were obtained with the same instrument
setup. Following \citet{mazeh02}, we observed the wavelength range
1.5515--1.5585 $\mu$m (central wavelength 1.555 $\mu$m; $H$ band)
at a resolving power of $R \approx 30000$ (slit size of $0\farcs35$). 
Exposure times were
varied between 1.0 and 3.3 hr by the queue observers based on sky
conditions. The typical exposure time was 2.6 hr (i.e.\ $\sim 0.01$
orbital phase; see \S\ref{lightcurve}), divided into 6 telescope nods
in an {\tt abbaab} pattern for the purposes of sky subtraction and
cosmic-ray rejection. An observation of the {\sc coravel-elodie}
high-precision radial-velocity standard \objectname{HD~50778}
\citep[K4 III;][]{udry99} was obtained immediately before or after
each observation of \bd\ to monitor the instrument wavelength zero
point. In addition, observations were obtained of a grid of late-type
spectral standards, with spectral types of M0 to M9, selected from
\citet{mazeh02} and \citet{mohanty03}. The standard-star observations
are summarized in Table~\ref{standards}.

The observations were processed using Interactive Data Language
procedures developed specifically for optimal extraction of {\it
Phoenix} spectra. The background subtraction step includes logic
to compensate for variations in the night sky lines due to grating
motion or brightness variations between nods. A bootstrap procedure
is used to determine a wavelength solution from ThNeAr calibration
spectra obtained immediately before and after the science exposures.
First an {\it a priori} wavelength dispersion is determined by matching a
{\it Phoenix} spectrum of a K giant with the KPNO FTS atlas of Arcturus
\citep{hinkle95}. Next, a small wavelength shift (less than 1 km s$^{-1}$)
is applied to match the mean wavelengths of the three strongest
ThNeAr lines as a function of position along the slit. Finally, a
multi-Gaussian fit is used to empirically determine the wavelengths
of the remaining ThNeAr lines. The individual exposures from the
different slit positions are combined and interpolated onto a uniform
dispersion and then continuum normalized using sixth-order polynomial
fits to the high points in the spectra. Typical signal-to-noise ratios
of the extracted spectra were $\sim 15$ per resolution element for
the nine observations of \bd\ and $\sim 50$ per resolution element
for the standard stars.

Absolute heliocentric radial velocities of the late-type standards
were determined relative to the {\sc coravel-elodie} standard star
HD~50778 (see above). To minimize the possibility of introducing
systematic velocity offsets due to large differences in spectral
type between this K4 standard and the late-M standards, we applied
velocity corrections in a step-wise fashion, determining the velocity
shift of each sequentially later type standard by cross-correlating
it with the standard from the previous step (e.g., M0 relative to K4,
M1 relative to M0, and so on). The radial velocities reported below
are thus formally on the {\sc coravel-elodie} system \citep{udry99}.

Radial velocities of the two components of \bd\ were determined
via the technique of Broadening Functions (BFs) described by
\citet{rucinski99}. As discussed in that study, BFs are less prone
than simple cross-correlation techniques to ``peak pulling," in which
closely spaced correlation features in double-lined binaries can
alter the positions of the correlation peak centroids, particularly
when the velocity separation of the components is on the order of
the spectral resolution. The BF analysis requires a radial-velocity
template spectrum that is well matched in spectral type to the target;
we determined that the M6.5 star LHS~292 from our grid of standards
was the best template for this purpose (see below).

In seven of the nine {\it Phoenix} spectra of \bd\ the BFs
show two clear peaks, an example of which is shown in Fig.\
\ref{bf-fig}. For these seven observations, radial velocities and their
uncertainties were determined using two-Gaussian fits to the BFs
\citep[e.g.][]{rucinski99}. We designate the component producing
the stronger peak in the BFs as the `primary' and the component
producing the weaker peak in the BFs the `secondary' (as we show
below, the primary component so defined is also the more massive
of the pair). One of the remaining two observations was by chance
taken during eclipse of the primary (orbital phase 0.75; see Fig.\
\ref{Ilc-fig}) and thus appears as a single-lined spectrum; the BF
peak was fit with a single Gaussian and its velocity assigned to the
secondary. The other observation occurs shortly after eclipse of the
secondary (orbital phase 0.12) and thus appears as a blended spectrum
producing a single, broadened BF peak; in this case two Gaussians
were fit, their initial centroids estimated from the preliminary orbit
solution of Paper~I and their widths fixed based on the widths of the
well-separated BF peaks from the other spectra. Radial velocities and
their uncertainties so measured are summarized in Table~\ref{rv-data}
and displayed in Fig.\ \ref{rv-fig}.

We found that the BF peaks of both components appeared strongest when
we used LHS~292, an old M6.5 dwarf (see Table~\ref{standards}),
as the radial-velocity template in the BF analysis. 
The BF peaks were $\sim 20\%$ weaker when
we used the M6 or M7 dwarfs as templates, and weaker still when we
used earlier or later templates.
This suggests that the spectral types of both
components of \bd\ are also $\sim$M6.5. 
Indeed, we found in Paper~I
(cf.\ Fig.\ 3 in that paper) that the isolated spectrum of the
primary component of \bd\ very closely matches that of LHS~292,
and that the $JHK$ colors
of \bd\ imply a spectral type in combined light of M6.5$\pm 0.5$
with negligible reddening. 
Taken together, the available evidence suggests that the components of
\bd\ have very similar spectral types of $\sim$M6.5$\pm 0.5$,
corresponding to $T_{\rm eff} \approx 2700$~K 
\citep{slesnick04,goli04}.

While our light-curve analysis below (\S\ref{analysis}) yields a
$T_{\rm eff}$ ratio of $\sim 1$, reinforcing the conclusion that
the components of \bd\ are very similar in spectral type,
we caution that our assignment of absolute spectral types is 
preliminary and subject to systematic uncertainty. 
The differences in spectral features seen in late-M standard stars
are extremely subtle for dwarfs with spectral types M4--M8
\citep[e.g.][]{bender05,prato02},
and the low S/N of our \bd\ spectra prohibits a detailed
analysis of spectral features.
Moreover, the low surface gravities and strong magnetic fields of 
young, low-mass objects
can bias classification of their spectra, in the sense that the true
spectral types may be 1--2 subtypes later than inferred from comparison
to old, high-gravity dwarfs \citep{mohanty03}. 
Thus, for our purposes here, we emphasize 
that the spectral match between \bd\ and the adopted radial-velocity
template is sufficiently good to permit precise radial velocity
measurements, and that the spectral types of the components of \bd\
are evidently very similar to one another.

Because the components of \bd\ have such similar spectral
types, the relative BF peak areas reflect the two components'
relative contributions to the total light of the system
\citep[e.g.][]{bayless06}. In particular, the primary component
apparently contributes approximately 50\% more flux than the
secondary at 1.555 $\mu$m (see Fig.\ \ref{bf-fig}). This flux
ratio provides an important additional constraint for removing
degeneracies in the determination of physical parameters, as we
discuss in \S\ref{analysis}.

The component masses that we determine in \S\ref{analysis}
follow directly from an orbit solution fit to the observed radial
velocities, and thus the uncertainties in the derived masses (and
in other properties that in turn depend on those masses) depend
sensitively on uncertainties in the radial velocity measurements. The
mean precisions of the radial-velocity measurements from the
BF analysis are formally 1.7 and 1.9 km~s$^{-1}$ for the \bd\
primary and secondary, respectively (see $\sigma_{\rm RV}$ column of
Table~\ref{rv-data}). However, the accuracy of these measurements
can potentially be degraded by various systematic effects. At
the instrument level, the stability of the wavelength zero-point
is of particular concern. To assess this, we cross-correlated the
first observation of the {\sc coravel-elodie} standard star HD~50778
against each subsequent observation of this star, providing a measure
of the instrument stability over the course of our {\it Phoenix}
observations. The resulting radial velocities of HD~50778 exhibit
an r.m.s.\ scatter of 0.45 km~s$^{-1}$ on the nights that \bd\
was observed, which is much smaller than the random errors in the
individual measurements of \bd. Another possible source of systematic
error is spectral-type mismatch between \bd\ and the radial-velocity
template. However, as discussed above, the radial-velocity template
that we used in the BF analysis appears to match the spectral types
of the \bd\ components very well. Moreover, the components of \bd\
have very similar spectral types to one another, so that any systematic
effects due to spectral mismatch with the template should affect both
components similarly, in which case only the center-of-mass velocity
will be affected. 

Finally, the observed radial velocities can be affected by spots
on the surfaces of the brown dwarfs that cause phase-dependent
distortions in the shapes of the spectral lines, and in fact we
show in \S\ref{spots} that spots are clearly present on one or
both of the components in \bd. While the data suggest that this
effect is small, we have not yet incorporated a physical spot
model into our analysis and thus any spot signatures remain as
potential systematics in the radial velocities. In V1174 Ori, for
example, a young eclipsing binary with photometric spot amplitudes
similar to those observed in \bd, we found that the resulting
radial-velocity distortions were as large as $\sim 1$ km~s$^{-1}$ at
certain orbital phases \citep{stass04}.  Analyses of other spotted,
low-mass eclipsing systems find similar effects; for example, GU Boo
shows distortions of $\sim 0.5$ km~s$^{-1}$ \citep{lopez05} and YY
Gem shows distortions of $\sim 1$ km~s$^{-1}$ \citep{torres02}. Any
spot-induced radial-velocity distortions in \bd\ would need to be
2--3 times larger than in these systems to be comparable to the
random measurement errors of $\sim 1.5$--2 km~s$^{-1}$ \citep[for
an example of such a system, see][]{neuhauser98}. Encouragingly,
the residuals of the radial-velocity measurements with respect to
the final orbit solution of \S\ref{analysis} ($\chi_\nu^2 = 0.7$
and 1.2 for the primary and secondary velocities, respectively; see
Table~\ref{rv-data}) do not indicate that systematic effects dominate
the radial-velocity measurement errors.

\subsection{Photometric observations: Light curves\label{lightcurve}}
We have continued to photometrically monitor \bd\ intensively with
the 1.0-m and 1.3-m telescopes at CTIO and, as of this writing, 
possess a total of 2404 $I_C$-band measurements spanning the time
period 1994 December to 2006 April. The observing campaign to date is
summarized in Table~\ref{obs-table}, and the individual measurements
are provided in Table~\ref{lc-data}. We have excluded a small number
of observations with photometric errors larger than 0.1 mag. Thus,
photometric errors on the individual measurements have a range of
0.01--0.1 mag, with a mean error of 0.024 mag and a median error of
0.020 mag.

A period search based on the phase dispersion minimization (PDM)
technique of \citet{stellingwerf78} reveals an unambiguous period of
$P = 9.7795557 \pm 0.000019$ d. The PDM technique is well suited to
periodic variability that is highly non-sinusoidal in nature, as is
the case for most detached eclipsing binaries. This updated period is
slightly shorter than, but not inconsistent with, the period reported
in Paper~I (see Table~\ref{properties-table}).

In Fig.~\ref{Ilc-fig} we show the $I_C$-band light curve of \bd\
folded on this period. Two distinct eclipses are clearly evident
and cleanly separated in phase, as is typical of fully detached
eclipsing binaries. One eclipse---the `primary eclipse' at orbital
phase $\sim 0.075$---is notably deeper than the other; this marks 
the time in the orbit when the hotter component is eclipsed which, 
as discussed in Paper~I and below in \S\ref{reversal}, in this 
system is the lower-mass component.

In addition, the time from primary to secondary eclipse is longer
than that from secondary to primary eclipse, indicating an eccentric
orbit. More quantitatively, the two parameters that determine the
shape and orientation of the orbit---the eccentricity, $e$, and the
argument of periastron, $\omega$---can be estimated from geometrical
considerations relating the orbital period, the durations of the
eclipses (6.8 hr and 10.1 hr, respectively), and their 6.5-day 
separation in time \citep[e.g.][]{kallrath99},
from which we estimate $e \approx 0.35$ and $\omega \approx 216^\circ$.
These initial estimates of $e$ and $\omega$ agree well with the more
precise values that we obtain from detailed light-curve modeling and
analysis below (\S\ref{analysis}; Table \ref{properties-table}).

In addition to the extensive $I_C$-band light curve measurements,
we have also now obtained a set of near-infrared ($JHK$) light curves.
These are presented and analyzed in a subsequent paper \citep{gomez06}, 
but we note here that the same features observed
in the $I_C$-band light curve are also seen at these other wavelengths. 

\subsection{Analysis\label{analysis}}
As in Paper~I, we have performed a simultaneous analysis of the
{\it Phoenix} radial velocities and the $I_C$-band light curve
using the eclipsing-binary algorithms of \citet[][updated 2005;
hereafter \model]{wd05} as implemented in the {\sc phoebe} code
of \citet{prsa05}. \model\ has become standard in the modeling
and analysis of all manner of eclipsing binary systems, having
grown in sophistication over the past 35 years to include, e.g.,
phase-dependent projection effects in eccentric orbits, radial-velocity
perturbations arising from proximity and eclipse effects, apsidal
motion, and reflection and limb-darkening effects. In its most advanced
implementation, the code can also treat the effects of surface spots,
model atmospheres, and asynchronous rotation, including any attendant
gravity brightening and radial-velocity perturbations arising from
non-sphericity and from phase-dependent variations in the shapes of
the components.

For the present study, we add information from the out-of-eclipse
variations in the light curve to model asynchronous rotation
of the components and as a first step toward accounting for the
presence of surface spots (\S\ref{spots}). As part of our ongoing
study of \bd, we plan to incorporate progressively more advanced
treatments, including a full investigation
of surface spots and inclusion of state-of-the-art brown-dwarf model
atmospheres. Here, as in Paper~I, we use simple blackbody spectra in
our light curve models. While brown-dwarf spectra are known to exhibit
strong departures from simple blackbodies, we show below that this
effect is unlikely to significantly alter our principal findings;
in particular, the reversal of temperatures with mass that we find
in \bd\ is probably not the result of improperly modeled spectra. On
the other hand, understanding the temperature reversal may ultimately
require a proper treatment of surface spots and their influence on
the time- and aspect-dependent emergent flux from the system. 

\subsubsection{Spot-related variations in the light curve\label{spots}}
To begin, we ran an initial \model\ fit to the light-curve and
radial-velocity data (Tables \ref{rv-data} and \ref{lc-data}) using
as initial values the system's orbital and physical parameters from
Paper~I, together with the updated orbital period from this study
(\S\ref{lightcurve}). The best-fit parameters and their formal
uncertainties are summarized in Table~\ref{properties-table}. The
reduced $\chi^2$ of this fit is large ($\chi_\nu^2=3.10$), and
the r.m.s.\ of the residuals relative to this fit (0.033 mag) is
accordingly larger than expected given the typical error on the
photometric measurements (\S\ref{lightcurve}). This suggests an
additional source of variability in the light curve---spots, for
example---not included in the model.

Surface spots are most commonly manifested as low-amplitude, periodic,
roughly sinusoidal variations superposed on the eclipse light curve
\citep[e.g.][]{stass04}. We searched for such low-amplitude, periodic
variations in the light curve of \bd\ by applying a Lomb-Scargle
periodogram \citep{scargle82} analysis to those portions of the
light-curve data obtained outside of eclipse. We performed this
periodogram analysis separately on each epoch of light-curve data (see
Table~\ref{obs-table}), as spot-related variations often show evolution
in amplitude and phase with time \citep{stass99,torres02,stass04}.

We find statistically significant periodic signals in all but the
first two epochs, these earlier epochs possessing relatively few
measurements. While the amplitudes of these periodic signals vary
slightly from epoch to epoch, they are always small ($\Delta m \lesssim
0.06$ mag, peak-to-peak). Importantly, the periods of these signals are
consistent across epochs, with a value of $P_{\rm spot} \approx 3.3$~d.
Spot-related variability observed in other eclipsing binaries is
similarly characterized by photometric amplitudes of a few percent and
periods that are constant over many epochs \citep[e.g.][]{ribas03},
most likely reflecting the rotation period of one or both binary
components. The implication here is that at least one of the brown
dwarfs in \bd\ is nearly always spotted, and rotates with $P_{\rm rot}
\approx 3.3$~d.

With a rotation period that is shorter than the orbital period by
a factor of (almost exactly) 3, the rotational and orbital angular
velocities of \bd\ have evidently not yet become pseudo-synchronized
with one another \citep[where the rotational angular velocity is
synchronized to the orbital angular velocity at periastron;][]{hut81},
which for the eccentricity of \bd\ would give $P_{\rm rot} / P_{\rm
orb} \approx 1/2$. Instead, the rotation of at least one component
in \bd\ is super-synchronous. Such super-synchronicity is probably
not surprising given the extreme youth of \bd\ (see below) and that
young brown dwarfs tend to be rapid rotators \citep{mohanty03}; tidal
effects have likely not yet had sufficient time to synchronize the
system \citep[e.g.][]{meibom05}.

A more detailed discussion of rotation in \bd\ is beyond the scope of
this paper. For our present purposes, we proceed to adopt the observed
$P_{\rm rot} = 3.3$~d in our final \model\ fit as a constraint on
the oblateness of both \bd\ components. In addition, we rectify the
$I_C$-band light curve by subtracting a sinusoid of $P = 3.3$~d, with
the amplitude and phase determined separately for each epoch of data,
as described above. The aim of this rectification procedure is to
``remove" the spot signal from the light curve and to thereby allow the
\model\ model to achieve goodness of fit without introducing additional
spot-fitting parameters that are poorly constrained at present. We
caution, however, that this procedure is {\it not} equivalent to
modeling the spots physically. The inclusion of additional light
curves at multiple wavelengths into our analyses \citep{gomez06}
will ultimately be required to constrain the physical properties of
the spots and to permit full modeling of their effects on the observed
light curves and radial velocities. Bearing these caveats in mind, we
proceed with our analysis below using the rectified $I_C$-band light
curve, and revisit the possible influence of spots in \S\ref{reversal}.

\subsubsection{Determination of physical parameters\label{finalfit}}
The final \model\ solution was determined by simultaneously fitting
the {\it Phoenix} radial velocities and the $I_C$-band light curve,
rectified as described above. The resulting orbital and physical
parameters of \bd\ are summarized in Table~\ref{properties-table},
the fit to the radial velocities shown in Fig.\ \ref{rv-fig}, 
the fit to the light curve displayed in Fig.\ \ref{Ilc-fig},
and the system geometry illustrated in Fig.\ \ref{orbit-fig}. 
With the light curve rectified, the
\model\ fit is now excellent; the r.m.s.\ of the residuals is 0.021
mag, comparable to the typical error on the photometric measurements
(\S\ref{lightcurve}), and the goodness-of-fit is correspondingly very
good, $\chi_\nu^2 = 1.035$.

A comparison of the orbital and physical properties of
\bd\ derived from the rectified and unrectified light curves
(Table~\ref{properties-table}) reveals mostly minor, statistically
insignificant, differences in the fit parameters. However, a few
parameters do show differences that are larger than their formal
uncertainties. For example, both $e$ and $\omega$ differ by $\sim
2\sigma$ (although these parameters are strongly correlated with one
another, so their uncertainties are not independent), suggesting
that the rectification procedure is not entirely free of small,
systematic couplings to some parameters. Thus, while a physical
treatment of spots in the \model\ model will ultimately be required
to achieve full accuracy in the determination of some parmaters, the
effects of spots on the most fundamental physical parameters that we
derive for \bd---the orbital parameters, in particular, and thus the
component masses and radii---are probably insignificant. Moreover,
if other parameters remain susceptible to spot-modeling effects,
these effects are evidently very subtle.

We note that there exists an alternative
\model\ solution to the one presented in Table~\ref{properties-table},
with nearly identical best-fit parameters except that the
component radii are reversed. Formally, the goodness-of-fit of this
alternative solution is inferior, with $\chi_\nu^2 = 1.062$, but
it is nonetheless acceptable. This particular type of degeneracy
is a general feature of eclipsing-binary solutions \citep[see,
e.g.,][]{stass04,lopez05}. Since only the component radii are
significantly different in the two solutions, the degeneracy can be
broken by adding the additional constraint of the luminosity ratio,
which is generally very different in the two solutions. In this
case, the preferred solution from Table~\ref{properties-table}
gives an $H$-band luminosity ratio of $L_1/L_2 = 1.63 \pm 0.13$,
whereas the alternative solution gives $L_1/L_2 = 0.69 \pm 0.06$. The
luminosity ratio predicted by the alternative solution---in which
the secondary is both warmer and larger, and thus more luminous,
than the primary---is clearly inconsistent with the ratio inferred
from the observed spectra, which imply $L_1/L_2 \sim 1.5$ in the $H$ 
band (see Fig.\ \ref{bf-fig}). 

\section{Results\label{results}}
\subsection{Masses and Radii}
With masses of $M_1 = 0.0569 \pm 0.0046\; {\rm M}_\odot$ and $M_2
= 0.0358 \pm 0.0028\; {\rm M}_\odot$, the components of \bd\ are
proven to be {\it bona fide} brown dwarfs. Moreover, with radii
of $R_1 = 0.674 \pm 0.023 \; {\rm R}_\odot$ and $R_2 = 0.485 \pm
0.018 \; {\rm R}_\odot$, these brown dwarfs are more comparable
in their physical dimensions to young, low-mass stars than to
old brown dwarfs, consistent with their being very young (see
below). Importantly, these mass and radius measurements are distance
independent. Moreover, systematic errors do not appear to dominate
over random measurement uncertainties (\S\ref{spectra}). Thus the
masses and radii of \bd\ reported here are accurate to 8\% and 3\%,
respectively. In \S\ref{tracks} we make use of these measurements for
an initial examination of the predictions of theoretical brown-dwarf
evolutionary models.

\subsection{Reversal of temperatures\label{reversal}}
The physical properties that we have determined for the brown dwarfs in
\bd\ are surprising in at least one important respect: The less-massive
secondary is hotter than the primary. Specifically, we find $T_{\rm
eff,2}/T_{\rm eff,1} = 1.064 \pm 0.004$ (Table~\ref{properties-table}),
which follows from the relative depths of the eclipses and bolometric
corrections from the model atmospheres (blackbodies in this case),
with small corrections for differences in occulted areas that occur due
to the mildly eccentric orbit (Fig.\ \ref{orbit-fig}). This finding
is highly statistically significant. Such a reversal of temperatures
with mass is not predicted by any theoretical model for coeval brown
dwarfs---in which temperature increases monotonically with mass---and
thus merits closer scrutiny. Here we consider the possible effects
of non-blackbody atmospheres and surface spots.

Brown dwarf spectra deviate significantly from an ideal blackbody,
due primarily to the strong wavelength dependence of molecular
opacity. The corresponding redistribution of flux into or out of any
specific wavelength depends on temperature, potentially affecting the
ratio of eclipse depths. However, the primary and secondary of \bd\
have very similar effective temperatures (\S\ref{spectra}), so both
components should have similar deviations from a blackbody.

To check this quantitatively, we have examined up-to-date synthetic
models of brown-dwarf spectra for temperatures appropriate to the
components of \bd. In particular, we use the solar-metallicity {\sc
cond}\footnote{At the temperatures of the components of \bd, the
{\sc dusty} atmosphere models of these authors are very similar and
would work equally well for our purposes. See also \citet{mohanty03}.}
atmosphere models of \citet{allard01}. As the masses and radii
of \bd\ imply $\log g \approx 3.5$ for both components, we use the
{\sc cond} models with this $\log g$ value also. The {\sc cond}
models incorporate recent opacities, including $\sim 500$ million
molecular lines, as well as dust formation and condensation at very
low temperatures. (For a more detailed discussion of these models, and
their applicability to young brown dwarfs in particular, we point the
reader to the extensive discussion in \citet{mohanty03}). The results
of our analysis are summarized in Fig.\ \ref{speccomp-fig}. We find
that, while dramatic departures from simple blackbodies are clearly
present in the model spectra, the ratio of surface brightnesses in the
$I$ band is consistent with that predicted from simple blackbodies
to within 2\%, which implies a $T_{\rm eff}$ ratio that is consistent
with the blackbody value to within 0.5\%.

As an additional check, we calculated the $T_{\rm eff}$'s of the
components of \bd\ independently at 1.25, 1.65, and 2.2 $\mu$m.
We use the ratio of surface fluxes at 0.8 $\mu$m (measured directly
from the ratio of eclipse depths in the $I_C$-band light curve;
\S\ref{analysis}) and the radii (measured directly from the eclipse
durations and the orbital velocities; Table~\ref{properties-table}),
together with an assumed distance to the Orion Nebula Cluster of 450
pc, empirical bolometric corrections from the literature
\citep{goli04}, and the observed $JHK$ magnitudes at these wavelengths
from \citet{carpenter01}: 
$K=13.58 \pm 0.02$, $J-H = 0.640 \pm 0.015$, and $H-K = 0.385 \pm 0.015$.
We derive temperatures of $T_1 = 2730$ K and
$T_2 = 2875$ K (1.25 $\mu$m), $T_1 = 2675$ K and $T_2 = 2860$ K (1.65
$\mu$m), and $T_1 = 2680$ K and $T_2 = 2795$ K (2.2 $\mu$m). In all
cases, the derived $T_{\rm eff}$'s are consistent with a spectral type
of $\sim$M6.5, and in all cases the reversal of temperatures persists.

Another potential contributor to non-blackbody spectra is the
presence of hot or cold spots (akin to solar plage or sunspots) on
the brown dwarfs, leading to spectra that derive from a combination
of temperatures. Cool spots have now been found to be present on many
brown dwarfs \citep[e.g.][]{scholz05} and may thus be expected on
one or both brown dwarfs in \bd\ as well. Indeed, we find a clear
periodic signal in the out-of-eclipse portions of the $I_C$-band
light curve, with a period of $P_{\rm spot} \approx 3.3$ d and a
semi-amplitude of $\sim 0.03$ mag (\S\ref{spots}); almost certainly
this is a rotationally modulated spot signal.

Such spot signals are commonly observed in the light curves of
close eclipsing binaries, though modeling these effects is often
a largely cosmetic exercise with little effect on the resulting
stellar properties. For example, in our analysis of the young
eclipsing binary \object{V1174 Ori} \citep{stass04}, we successfully
reproduced the observed out-of-eclipse light-curve variations by
including spots in the light-curve model, but the derived system
parameters were altered only negligibly as a result \citep[see
also][]{torres02,ribas03,lopez05}.

On the other hand, accurate modeling of spot effects can be crucially
important for the proper interpretation of certain systems. W~UMa
stars of the so-called ``W type" are a particularly good case
in point. A defining characteristic of these systems is that the
deeper eclipse corresponds to the occultation of the physically
smaller and lower-mass secondary, similar to what is observed in
\bd. One interpretation advanced early on by \citet{rucinski74}
is that the secondary is hotter than the primary (typically by
$\sim 5\%$), perhaps due to thermal effects arising from mass- and
energy-transfer in these contact systems. However, an alternate
interpretation \citep[e.g.][]{eaton80} is that cool spots on the
primary, if preferentially located along the the eclipsed latitude and
more-or-less uniformly distributed in longitude, can have the effect
of lowering the surface brightness of the eclipsed regions on the
primary during transit by the secondary, thus making the eclipse of the
primary shallower. Such a model is generally very
delicate in the details, and requires careful arrangement of the
putative spots in order that they produce only mild ($\lesssim 0.05$
mag) out-of-eclipse variations in the light curve; the inferred spot
configurations can resemble a ``leopard print" pattern in some cases
\citep[e.g.][]{linnell91}. 

We are now experimenting with more sophisticated light-curve models
that include the effects of spots. 
In the meantime, we emphasize here that the ratio of eclipsed surface
brightnesses implied by our current light-curve model is robust;
the higher-mass primary really does radiate less per unit area
than the secondary, at least at those surface elements that are
eclipsed by the secondary. Thus, in what follows, we continue to
explore the implications of a temperature reversal in \bd. 

\subsection{Physical association with the Orion
Nebula star-forming region: Evidence for youth}
Strong evidence for the physical association of \bd\ with the
Orion Nebula Cluster, and hence of its youth, is provided by its
center-of-mass velocity. The observed value of $\gamma = 24.1
\pm 0.4$ km s$^{-1}$ (Table~\ref{properties-table}) is within
1 km s$^{-1}$ of the systemic radial velocity ($25 \pm 1.5$ km
s$^{-1}$) of kinematic members of this active star-forming region
\citep{stass99,sicilia-aguilar05}.

In addition, we can derive a distance to \bd\ by comparing
the total system luminosity to its observed flux. To
calculate the luminosity, we use the directly measured radii
(Table~\ref{properties-table}) together with the effective
temperatures and apply the Stefan-Boltzmann relation, $L=4\pi R^2
\sigma_B T_{\rm eff}^4$. The adopted spectral type of M6.5$\pm$0.5
for the \bd\ primary (\S\ref{spectra}) implies $T_{\rm eff,1} =
2715 \pm 100$~K based on recent calibrations of the $T_{\rm eff}$
scale for brown dwarfs \citep{mohanty03,slesnick04,goli04}, though
as mentioned earlier, systematic uncertainties in the $T_{\rm
eff}$ scale may be as large as $\sim 200$~K. From the measured
temperature ratio of $T_{\rm eff,2}/T_{\rm eff,1} = 1.064 \pm
0.004$ (Table~\ref{properties-table}), we obtain $T_{\rm eff,2} =
2820 \pm 105$~K. This then gives component luminosities of $L_1 =
0.0223 \pm 0.0034\; {\rm L}_\odot$ and $L_2 = 0.0148 \pm 0.0023\;
{\rm L}_\odot$, and a total system luminosity of $L = 0.0372 \pm
0.0060\; {\rm L}_\odot$. 
Adopting an apparent magnitude of $m_K=13.58\pm 0.02$ \citep{carpenter01}
and bolometric corrections appropriate for the observed $T_{\rm eff}$'s
\citep{goli04} yields a derived distance to \bd\ of $456 \pm 34$
pc, assuming no extinction.  
This distance determination is consistent with
the distance to the Orion Nebula of $480 \pm 80$ pc \citep{genzel89}.

The Orion Nebula Cluster is extremely young, with an age that has been
estimated to be just $1^{+2}_{-1}$ Myr \citep{palla99,hill97}. With
a center-of-mass velocity and distance that are both consistent with
membership in this cluster, \bd\ is probably also very young, likely
having formed within the past few Myr. In addition, as discussed
in Paper~I, the $JHK$ colors of \bd\ place an upper limit on the
extinction of $A_V< 0.75$, and thus limit the amount of remnant
material available to the brown dwarfs for ongoing accretion.
Given the eclipsing nature of the system, any disk material would
necessarily be seen edge-on and would thus produce a large amount of
extinction and reddening. Thus, while the colors of \bd\ are formally
consistent with a small amount of interstellar extinction and/or a
small amount of remnant disk material, the currently observed masses
are unlikely to change significantly over time. These brown dwarfs
will likely forever remain brown dwarfs.

\subsection{Comparison to theoretical brown-dwarf evolutionary tracks
\label{tracks}}
While we expect that our ongoing analysis of \bd\ will further improve
the accuracy of the measured system parameters, the present accuracy is
sufficient to permit an initial examination of theoretical brown-dwarf
evolutionary models. Here we consider two of the more widely used
sets of tracks: those of \citet{baraffe98}\footnote{We use the tracks
with convection mixing-length parameter $\alpha=1.0$, as recent
analyses of young, low-mass stars and brown dwarfs suggest that such
low-efficiency convection best matches the observed physical properties
of these objects \citep[e.g.][]{stass04,mathieu06,torres06}.} and
those of \citet[][updated 1998]{dantona98}. These models differ in
their treatment of brown-dwarf atmospheres and of energy transport
(convection) in brown-dwarf interiors, as well as in the choice of
initial conditions \citep[for a more in-depth discussion of differences
in these models, see][]{siess00,baraffe02}.

Fig.\ \ref{models-fig} compares the observed radii, effective
temperatures, and luminosities of \bd\ with the values theoretically
predicted by these models for brown dwarfs with masses of $M_1 =
0.0569 \pm 0.0046\; {\rm M}_\odot$ and $M_2 = 0.0358 \pm 0.0028\;
{\rm M}_\odot$ (Table~\ref{properties-table}). 

We note first that, generally speaking, the agreement between the
observed and theoretically predicted properties of \bd\ is quite
good. The models predict that these very young brown dwarfs should,
at an age of $\sim 1$ Myr, be significantly larger, warmer, and more
luminous than their older counterparts---and that is in fact what
we see. Indeed, between 1 and 100 Myr, a brown dwarf with a mass of
0.057 ${\rm M}_\odot$ (the mass of the \bd\ primary) is predicted
to shrink by 500\%, cool by several hundred K (or, equivalently,
go from an M spectral class to an L spectral class), and dim by 1.5
orders of magnitude (Fig.\ \ref{models-fig}). The recently measured
radius ($R = 0.12\; {\rm R}_\odot$) of the old and low-mass ($M =
0.09\; {\rm M}_\odot$) star in the eclipsing binary system OGLE-TR-122
\citep{pont05} confirms that stars with near-brown-dwarf masses have
very small radii ($\sim 1$ R$_{\rm Jup}$) when they are old. Thus,
the fact that \bd\ comprises young brown dwarfs that are both large
and luminous---and even simply that they are of M spectral class,
as predicted---is a testament to the generally good predictive power
of current theoretical models of brown dwarfs.

At a more detailed level, the models of \citet{dantona98} predict radii
and luminosities that are more consistent with the observed values. The
difference is most pronounced in the radii; the \citet{baraffe98}
models under-predict both radii by $\sim 10\%$. This finding of
under-predicted model radii is similar in sense and magnitude to that
found from recent efforts to derive the physical properties of low-mass
stars and brown dwarfs through detailed modeling of their spectra
\citep{mohanty04a,mohanty04b}. In addition, while both sets of models
perform reasonably well with respect to the observed luminosities,
the agreement is somewhat better for the \citet{dantona98} models;
it appears possible that more accurate measurements will reveal an
under-prediction of luminosity by the \citet{baraffe98} models for
the lower-mass secondary.

But perhaps more importantly, the reversal of temperatures in \bd\
remains puzzling; the relationship between effective temperature and
mass is predicted by both sets of models to be monotonic for brown
dwarfs of the same age. However, this expectation disappears for two
brown dwarfs of differing age. The observed effective temperatures can
be seen as consistent with the \citet{dantona98} models if the primary
is taken as being modestly younger than the secondary ($\Delta\tau
\approx 0.5$ Myr). Indeed, the temperatures, radii, and luminosities of
\bd\ all remain in marginally good agreement with these models for such
an age difference, at a mean age of 1 Myr (Fig.\ \ref{models2-fig}),
if we also adjust the observed $T_{\rm eff}$ scale cooler by 70 K.
Such a shift is well within the current systematic uncertainty of the
brown-dwarf $T_{\rm eff}$ scale (\S\ref{spectra}). Thus, a question
raised by our findings is whether current brown-dwarf formation theory
can accommodate a scenario in which two brown dwarfs---that are part of
the same, very young, binary system---can have sufficiently different
ages to allow for reversed temperatures as we have observed in \bd.

\section{Discussion\label{discussion}}

\subsection{\bd\ as a case study in dynamical brown-dwarf formation?}
In truth, we have very little understanding about how binaries
with components as close as those in \bd\ are formed \citep[see,
e.g.,][]{bonnell01,tohline02}. (\bd\ is, after PPl~15 in the Pleiades
\citep{basri99}, the shortest-period brown-dwarf binary system
yet discovered). With a few exceptions \citep[e.g.][]{tohline01},
fission seems to be ruled out. {\it In situ} mechanisms involving
dynamic cloud fragmentation and disk fragmentation have not yet proven
successful in creating such close binaries, but remain to be fully
developed. In particular, the role of orbital migration in the presence
of a massive circumbinary disk needs to be considered. Certainly,
if the two brown dwarfs formed together as a close binary, we have
essentially no {\it a priori} expectations for whether they should
be observed to be coeval to within 0.5 Myr.

Alternatively, recent theoretical work \citep{reipurth01}, as well
as detailed numerical simulations \citep{bate02a,bate02b,bate05},
suggest that dynamical interactions may be integral to the
formation of brown dwarfs. The argument is essentially that strong
gravitational interactions in multiple-body encounters provide
a feasible mechanism for disrupting the accretion process, and
thereby preventing the accumulation of mass by objects that would
otherwise have become stars. This hypothesis remains under debate
\citep[e.g.][]{maxted05}. Still, it is tempting to speculate that
the components of \bd\ did not form together as a binary, but rather
formed separately---with the primary forming later---and then were
later married through a dynamical interaction. In such a scenario, it
is possible that the resulting binary system---comprising two objects
that were not originally formed together---may exhibit seemingly
peculiar characteristics, such as the observed reversal of effective
temperatures, that in fact reveal the non-coeval nature of the system.

There are at least two specific scenarios involving multiple-body
interactions that might pertain to the origin of \bd. One involves
a low-mass binary pair that interacts with a more massive third body
from elsewhere in the cluster. Simulations show that, if the binary
is not simply broken apart by the encounter, the lower-mass member
of the binary is ejected and replaced by the massive incoming object
\citep{bate02a}. The resulting binary would then likely consist of
non-coeval members. A serious concern with this scenario in the context
of \bd\ is that it is unlikely in three-body encounters within clusters
that the most massive object would have a mass of only 0.06 M$_\odot$.

Thus, a second scenario is that several objects formed in the
fragmentation of a small molecular core, forming an unstable multiple
system. The consequent rapid dynamical evolution of the system may have
both terminated accretion and formed a hard binary \citep{bate02b}. In
such a fragmentation scenario the relative core-collapse times and
evolutionary zero points might easily vary by 0.5 Myr, of order the
age difference needed to explain the temperature reversal of \bd\
in the context of some evolutionary models (Fig.\ \ref{models2-fig}).

The position of \bd\ might be taken as further evidence that
dynamics have been important in its history. With a projected
separation of 2.8 pc from the center of the Orion Nebula, \bd\
is situated more than 10 core radii from the center of the
Orion Nebula Cluster \citep{hillenbrand98}; perhaps the pair was
ejected from the cluster center during a dynamical encounter. The
one-dimensional velocity dispersion in the cluster is $\sim 2$ km
s$^{-1}$ \citep{vanaltena88,hillenbrand98,stass99,sicilia-aguilar05},
so to reach its current position in 1 Myr, if ejected from the cluster
core, \bd\ would need a somewhat higher-than-usual velocity of $\sim 3$
km s$^{-1}$ in the plane of the sky. However,
the measured center-of-mass radial velocity of \bd\
does not deviate significantly from the cluster velocity.
An alternative
interpretation for the position of \bd\ at the outskirts of the
cluster may be that it is not a member of the Cluster proper, but
rather a member of the more widely distributed population of young
stars in the region surrounding the Orion Nebula \citep{warren78}. The
binary's proper motion is the critical measurement needed to establish
whether \bd\ was ejected from the cluster center.

\subsection{Missing physics in theoretical brown-dwarf evolutionary
tracks?} 
An alternative explanation is that the evolutionary tracks are
deficient with respect to some critical physical ingredient(s). For
example, the presence of strong magnetic fields on one or both brown
dwarfs in \bd\ could be affecting energy transport and thereby altering
their physical structure.

Indeed, recent analyses of several young, low-mass eclipsing binaries
\citep[e.g.][]{stass04,covino04,torres06} indicate systematic
discrepancies in the models. In particular, the observed effective
temperatures are cooler than expected, and the observed radii larger
than expected. These discrepancies are especially pronounced among
the most magnetically active stars (as traced by X-ray emission and
other proxies). One possible interpretation is that strong magnetic
fields inhibit energy transport in these otherwise fully convective
stars, resulting in a decrease of the surface temperature and a
corresponding increase in radius so as to radiate the same total flux
\citep{montalban06}. Such an interpretation would be consistent
with, and would help explain, the emerging observational evidence
for suppressed convection in young, low-mass stars \citep[e.g.][and
references therein]{stass04,mathieu06,torres06}.

In this context, the temperature reversal in \bd\ might be taken as
evidence for a strong magnetic field on the higher-mass primary that
is causing a sufficient decrease in its surface temperature to make
it effectively cooler than the lower-mass secondary. By inference,
the secondary would be interpreted as being less magnetically
active. The observational evidence is strong for magnetic activity
in young, low-mass stars and brown dwarfs of early- and mid-M type
\citep[e.g.][]{mohanty02,stass04b,preibisch05}. Moreover, the evidence
in fact shows a marked decline in magnetic activity at very late M
types \citep[e.g.][]{gizis00,mohanty02}, suggesting that brown dwarfs
with roughly the mass of the \bd\ secondary and below are not capable
of generating strong fields. The idea of a magnetically active primary
would also be consistent with the primary being heavily spotted (see
\S\ref{reversal}). 

If we are to explain the anomalously low effective temperature of
the \bd\ primary in this way, we should then also expect its radius
to be larger than theoretically predicted, as the one effect goes
hand in hand with the other \citep{montalban06}. As discussed above
(\S\ref{tracks}), whether the \bd\ radii agree with theoretical
predictions depends on one's choice of model. The \citet{baraffe98}
tracks do indeed suggest over-sized radii in \bd\ (for {\it both}
components), though this may simply reflect the truncation of
those tracks at 1 Myr. On the other hand, the \citet{dantona98} tracks
agree with the observed radii very well (Fig.\ \ref{models-fig}),
with no need to invoke missing physics. 

\section{Summary and Conclusions\label{conclusions}}
\bd\ is the first known eclipsing binary system comprising two
brown dwarfs. Satisfying both kinematic and distance requirements
for physical association with the young ($\sim 1$ Myr) Orion Nebula
Cluster, \bd\ provides the only direct, accurate measurements of
the fundamental physical properties of newly formed sub-stellar
objects. The masses that we measure are accurate to $\sim 10\%$,
the radii accurate to $\sim 5\%$, the ratio of effective temperatures
accurate to $\sim 1\%$, and all are distance independent. As such,
\bd\ represents an important benchmark for theoretical models of
brown-dwarf formation and evolution.

Encouragingly, we find that current brown-dwarf evolutionary tracks
are, broadly speaking, successful in predicting the fundamental
physical properties of these young brown dwarfs. More quantitatively,
of the two sets of theoretical models considered here, we find that
the models of \citet{dantona98} yield mass-radius and mass-luminosity
relationships that best agree with the empirically determined ones. The
models of \citet{baraffe98} predict radii and luminosities that are
1.5--2$\sigma$ smaller than the observed values.

However, the reversal of component effective temperatures with mass
in \bd\ is unexpected and unexplained. We have considered here two
possible interpretations of this intriguing result. The first is that
the components of \bd\ are mildly non-coeval, with the higher-mass
primary being $\sim 0.5$ Myr younger than the secondary. The models of
\citet{dantona98} are in fact consistent with the observed temperature
reversal for such an age difference. A second hypothesis is that
strong magnetic activity on the primary is inhibiting convection,
and thereby lowering its surface temperature.

Neither of these interpretations is wholly satisfying, and neither
is obviously discreditable. Binary formation theory is largely
silent on the subject of coevality at the level of $\sim 0.5$ Myr,
and theorists have long warned the star-formation community about
the limited applicability of evolutionary track chronometry at such
early ages (the model zero-points being arbitrary in most cases). In
addition, while the observational evidence is strong that young brown
dwarfs can be magnetically active, the effects of magnetic fields
on brown-dwarf structure and evolution have yet to be consistently
modeled or fully understood.

\acknowledgments
This research is supported by NSF grant AST-0349075, and by a Cottrell
Scholar award from the Research Corporation, to K.G.S., and by NSF
grant AST-9731302 to R.D.M.
This work is based in part on observations obtained through queue
program GS-2002B-Q-11 at the Gemini Observatory, which is operated
by the Association of Universities for Research in Astronomy, Inc.,
under a cooperative agreement with the NSF on behalf of the Gemini
partnership: the National Science Foundation (United States), the
Particle Physics and Astronomy Research Council (United Kingdom),
the National Research Council (Canada), CONICYT (Chile), the
Australian Research Council (Australia), CNPq (Brazil) and CONICET
(Argentina). 
The WIYN Observatory is a joint facility of the University of
Wisconsin--Madison, Indiana University, Yale University, and the
National Optical Astronomy Observatories.

\clearpage

\begin{deluxetable}{lcrc}
\tablecolumns{4}
\tablewidth{0pt}
\tablecaption{Standard stars\label{standards}}
\tablehead{
\colhead{Name} & \colhead{SpTy} & \colhead{Exp.\ time (s)} & \colhead{Ref.}
}
\startdata
GJ~328 & M0 & 180 & \citet{mazeh02} \\
GJ~382 & M1.5 & 180 & \citet{mazeh02} \\
GJ~447 & M4 & 120 & \citet{mohanty03} \\
GJ~406 & M6 & 120 & \citet{mohanty03} \\
LHS~292 & M6.5 & 3600 & \citet{mazeh02} \\
LHS~3003 & M7 & 900 & \citet{mohanty03} \\
GJ~3655 & M8 & 3600 & \citet{mohanty03} \\
BRIB 1507$-$0229 & M9 & 3600 & \citet{mohanty03} \\
\enddata
\end{deluxetable}

\clearpage

\begin{deluxetable}{rrrrr}
\tablecolumns{5}
\tablewidth{0pt}
\tablecaption{Radial velocity measurements of \bd\label{rv-data}}
\tablehead{
\colhead{HJD\tablenotemark{a}} & \colhead{Phase\tablenotemark{b}} &
\colhead{R.V.\tablenotemark{c}} & \colhead{$\sigma_{\rm RV}$} &
\colhead{$(O-C)$\tablenotemark{d}} \\
\colhead{} & \colhead{} & \colhead{km s$^{-1}$} & \colhead{km s$^{-1}$} &
\colhead{km s$^{-1}$}
}
\startdata
\cutinhead{Primary}
  2452623.74805  &    0.75  &  \nodata &  \nodata &  \nodata \\
  2452624.75701  &    0.85  &    9.59  &    1.41  &    1.34 \\ 
  2452625.72555  &    0.95  &    2.00  &    2.22  &    1.32 \\ 
  2452626.65154  &    0.04  &    9.84  &    1.75  &   $-$2.79 \\ 
  2452649.72938  &    0.40  &   38.02  &    1.41  &    1.30 \\ 
  2452650.67281  &    0.50  &   34.21  &    1.14  &    0.31 \\ 
  2452655.74512  &    0.02  &    6.61  &    2.56  &   $-$0.98 \\ 
  2452656.70824  &    0.12  &   25.58  &    1.28  &   $-$0.84 \\ 
  2453432.57920  &    0.45  &   37.09  &    2.11  &    1.62 \\
                 &  &  & \multicolumn{2}{c}{$\chi_\nu^2 = 0.7$} \\
\cutinhead{Secondary}
  2452623.74805  &    0.75  &   29.69  &    1.17  &   $-$2.50 \\ 
  2452624.75701  &    0.85  &   48.53  &    1.73  &   $-$0.41 \\ 
  2452625.72555  &    0.95  &   61.35  &    2.07  &    0.40 \\ 
  2452626.65154  &    0.04  &   42.43  &    2.97  &    0.43 \\ 
  2452649.72938  &    0.40  &    4.18  &    1.24  &    0.37 \\ 
  2452650.67281  &    0.50  &    5.07  &    2.00  &   $-$3.21 \\ 
  2452655.74512  &    0.02  &   50.04  &    2.35  &    0.05 \\ 
  2452656.70824  &    0.12  &   22.58  &    1.31  &    2.45 \\ 
  2453432.57920  &    0.45  &    6.64  &    2.64  &    0.85 \\
                 & &  & \multicolumn{2}{c}{$\chi_\nu^2 = 1.2$} \\
\enddata
\tablenotetext{a}{Heliocentric Julian Date.}
\tablenotetext{b}{Orbital phase, relative to ephemeris of
Table~\ref{properties-table}.}
\tablenotetext{c}{Heliocentric radial velocity.}
\tablenotetext{d}{Residual relative to \model\ solution (see text).}
\end{deluxetable}

\clearpage

\begin{deluxetable}{rccr}
\tablecolumns{4}
\tablewidth{0pt}
\tablecaption{Summary of $I_C$-band time-series photometry\label{obs-table}}
\tablehead{
\colhead{Obs.} & \colhead{Telescope} & 
\colhead{HJD range\tablenotemark{a}} & 
\colhead{$N_{\rm obs}$\tablenotemark{b}}
}
\startdata
    1   &  KPNO 0.9-m &  49699.70   --    49703.93 &  40 \\
    2   &  WISE 1.0-m &  49698.35   --    49714.42 &  47 \\
    3   &  WIYN 0.9-m &  52227.78   --    52238.01 & 102 \\
    4   &  WIYN 0.9-m &  52595.75   --    52624.95 & 144 \\
    5   &  SMARTS 0.9-m &  52622.57   --    52631.51 & 122 \\
    6   &  SMARTS 1.3-m &  52922.73   --    53081.57 & 347 \\
    7   &  SMARTS 0.9-m &  53011.57   --    53024.77 & 205 \\
    8   &  SMARTS 1.3-m &  53280.74   --    53340.73 & 230 \\
    9   &  SMARTS 1.0-m &  53373.56   --    53386.79 & 184 \\
   10   &  SMARTS 1.3-m &  53403.53   --    53445.60 & 338 \\
   11   &  SMARTS 1.3-m &  53646.82   --    53727.69 & 204 \\
   12   &  SMARTS 1.0-m &  53719.59   --    53727.83 & 190 \\
   13   &  SMARTS 1.3-m &  53745.64   --    53846.48 & 251 \\
\enddata
\tablenotetext{a}{Range of Heliocentric Julian Dates ($2400000+$).}
\tablenotetext{b}{Number of observations.}
\end{deluxetable}

\clearpage

\begin{deluxetable}{rrrr}
\tablecolumns{4}
\tablewidth{0pt}
\tablecaption{Differential $I_C$-band light curve of \bd\label{lc-data}}
\tablehead{
\colhead{HJD\tablenotemark{a}} & \colhead{$\Delta m$\tablenotemark{b}} & 
\colhead{$\sigma_m$} & \colhead{Obs.\tablenotemark{c}}
}
\startdata
  2453686.70525 &   0.006 &   0.020 &  11 \\
  2453686.71293 &   0.016 &   0.054 &  11 \\
  2453687.76841 &  $-$0.004 &   0.052 &  11 \\
  2453687.77617 &  $-$0.006 &   0.052 &  11 \\
  2453687.78389 &   0.020 &   0.029 &  11 \\
  2453687.79158 &  $-$0.002 &   0.029 &  11 \\
  2453688.73523 &  $-$0.001 &   0.020 &  11 \\
  2453688.74295 &  $-$0.008 &   0.061 &  11 \\
  2453688.75066 &   0.022 &   0.020 &  11 \\
  2453688.75835 &   0.030 &   0.020 &  11 \\
  2453689.77018 &   0.262 &   0.025 &  11 \\
  2453689.77783 &   0.220 &   0.020 &  11 \\
  2453689.79309 &   0.237 &   0.062 &  11 \\
  2453690.71340 &   0.008 &   0.020 &  11 \\
\enddata
\tablenotetext{a}{Heliocentric Julian Date}
\tablenotetext{b}{Differential $I_C$ magnitude (arbitrary zero-point).}
\tablenotetext{c}{Source of measurement (see Table~\ref{obs-table}).}
\tablecomments{The full table is available in the electronic version of
the Journal. A portion is shown here for guidance regarding its form
and content.}
\end{deluxetable}

\clearpage

\begin{deluxetable}{lrrr}
\rotate
\tablecolumns{4}
\tablewidth{0pt}
\tablecaption{Orbital and physical parameters of 
              \bd\label{properties-table}}
\tablehead{ 
\colhead{} & \colhead{Paper~I} & \multicolumn{2}{c}{This study} \\
\colhead{} & \colhead{} & \colhead{Unrectified\tablenotemark{a}} & \colhead{Rectified\tablenotemark{a}}
}
\startdata
Orbital period, $P$ [d] & $9.779621 \pm 0.000042$ & \multicolumn{2}{c}{$9.779556 \pm 0.000019$\tablenotemark{b}} \\
Time of periastron (Bessellian year), $T_0$ & $2001.863650 \pm 0.000095$ & $2001.863903 \pm 0.000160$ & $2001.863765 \pm 0.000071$ \\
Eccentricity, $e$ & $0.3225 \pm 0.0060$ & $0.3354 \pm 0.0049$ & $0.3276 \pm 0.0033$ \\
Orientation of periastron, $\omega$ [$^\circ$] & $215.4 \pm 1.1$ & $219.2 \pm 1.4$ & $217.0 \pm 0.9$ \\
Semi-major axis, $a \sin i$ [{\sc a.u.}] & $0.0398 \pm 0.0010$ & $0.0406 \pm 0.0016$ & $0.0406 \pm 0.0010$ \\
Center-of-mass velocity, $\gamma$ [km s$^{-1}$] & $24.1 \pm 0.4$ & $24.1 \pm 0.4$ & $24.1 \pm 0.4$ \\
Mass ratio, $q \equiv M_2/M_1$ &  $0.625 \pm 0.018$ & $0.622 \pm 0.022$ & $0.631 \pm 0.015$ \\
Total mass, $(M_1 + M_2) \sin^3 i$ [M$_\odot$] & $0.0880 \pm 0.0076$ & $0.0932 \pm 0.0111$ & $0.0932 \pm 0.0073$ \\
Inclination, $i$ [$^\circ$] & $88.8 \pm 0.2$ & $89.4 \pm 0.3$ & $89.2 \pm 0.2$  \\
Primary semi-amplitude, $K_1$ [km s$^{-1}$] & \nodata & $18.37 \pm 1.01$ & $18.49 \pm 0.67$ \\
Secondary semi-amplitude, $K_2$ [km s$^{-1}$] & \nodata & $29.55 \pm 1.24$ & $29.30 \pm 0.81$ \\
Primary mass, $M_1$ [M$_\odot$] & $0.0541 \pm 0.0046$ & $0.0575 \pm 0.0069$ & $0.0572 \pm 0.0045$ \\
Secondary mass, $M_2$ [M$_\odot$] & $0.0340 \pm 0.0027$ & $0.0358 \pm 0.0043$ & $0.0360 \pm 0.0028$ \\
Primary radius, $R_1$ [R$_\odot$] & $0.669 \pm 0.034$ & $0.673 \pm 0.037$ & $0.675 \pm 0.023$  \\
Secondary radius, $R_2$ [R$_\odot$] & $0.511 \pm 0.026$ & $0.485 \pm 0.029$ & $0.486 \pm 0.018$ \\
Primary gravity, $\log g_1$ & \nodata & $3.62 \pm 0.14$  & $3.62 \pm 0.10$ \\
Secondary gravity, $\log g_2$ & \nodata & $3.54 \pm 0.14$  & $3.54 \pm 0.09$ \\
Effective temperature ratio, $T_2/T_1$ &  $1.054 \pm 0.006$ & $1.062 \pm 0.006$ & $1.064 \pm 0.004$ \\
\enddata
\tablenotetext{a}{\model\ solutions based on fits to the unrectified
and rectified $I_C$-band light curve (see \S\ref{analysis}).}
\tablenotetext{b}{Uncertainty in the period is from a phase dispersion
minimization \citep{stellingwerf78} analysis on the $I_C$-band light curve
(see \S\ref{lightcurve}); the period is held fixed in the \model\ fit
and its uncertainty propagated into the uncertainties of derived
quantities.}
\end{deluxetable}

\clearpage

\figcaption[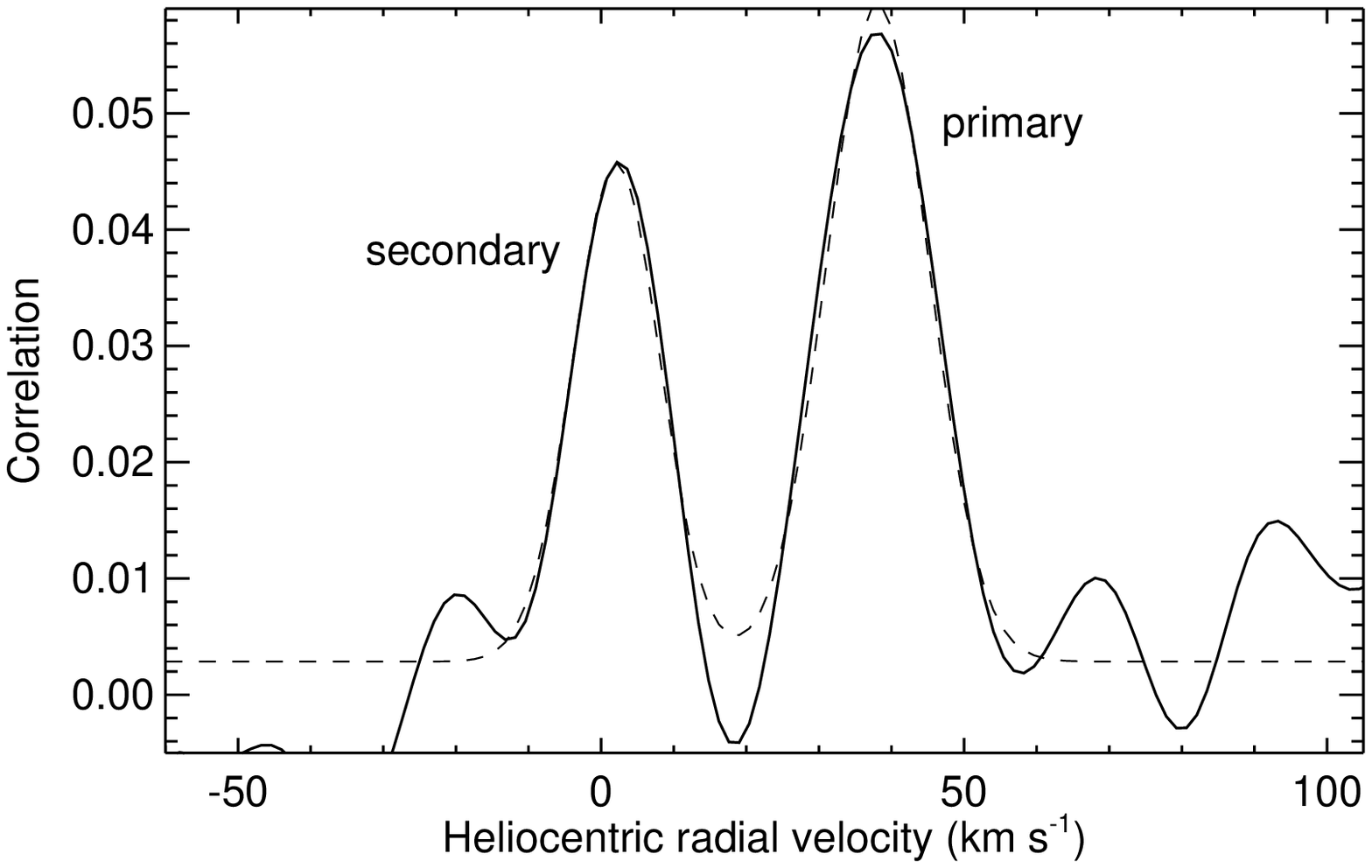]{\label{bf-fig}
Determination of the radial velocities of the primary and secondary
components of \bd\ using the Broadening Function (BF) analysis of
\citet{rucinski99}. The example BF shown here (solid line) is from the
spectrum of HJD 2452649 (see Table~\ref{rv-data}), where the primary
and secondary are at near maximum velocity separation. An M6.5 dwarf
was used as the radial-velocity template (see \S\ref{spectra}). The
dashed line shows a two-gaussian fit to the BF, from which velocity
centroids and their formal uncertainties are determined. The ratio
of the peak areas from this fit is 1.6, indicating that the primary
dominates the light of the system, contributing $\sim 60\%$ more
flux than the secondary at the wavelength of these observations
(1.555 $\mu$m). }


\figcaption[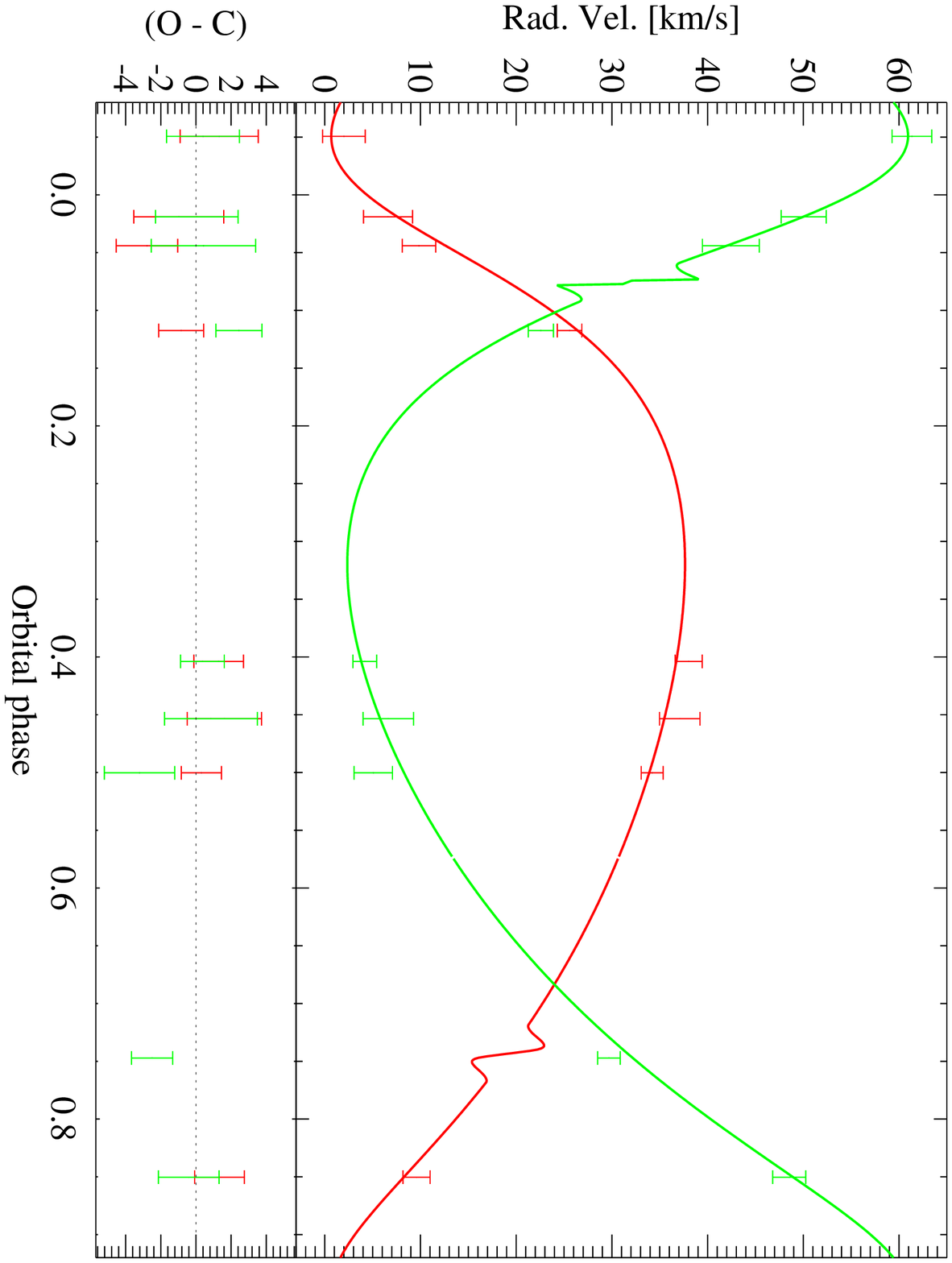]{
\label{rv-fig}
Radial velocity measurements of \bd. The individual radial
velocity measurements from Table~\ref{rv-data} are plotted 
(primary measurements in
green, secondary measurements in red), folded on the orbital period and
phased to the time of periastron (see Table~\ref{properties-table}).
The solid lines are \model\ models based on a simultaneous fit to the
radial velocity measurements and the rectified $I_C$-band light curve
(see Table~\ref{lc-data} and Fig.\ \ref{Ilc-fig}). Distortions in the
model curves near phases 0.075 and 0.75 are due to the brief occultations 
of the approaching and receding limbs of each component when it is 
eclipsed. Residuals are shown at bottom. Note: This figure appears in
color in the electronic version of the journal.
}


\figcaption[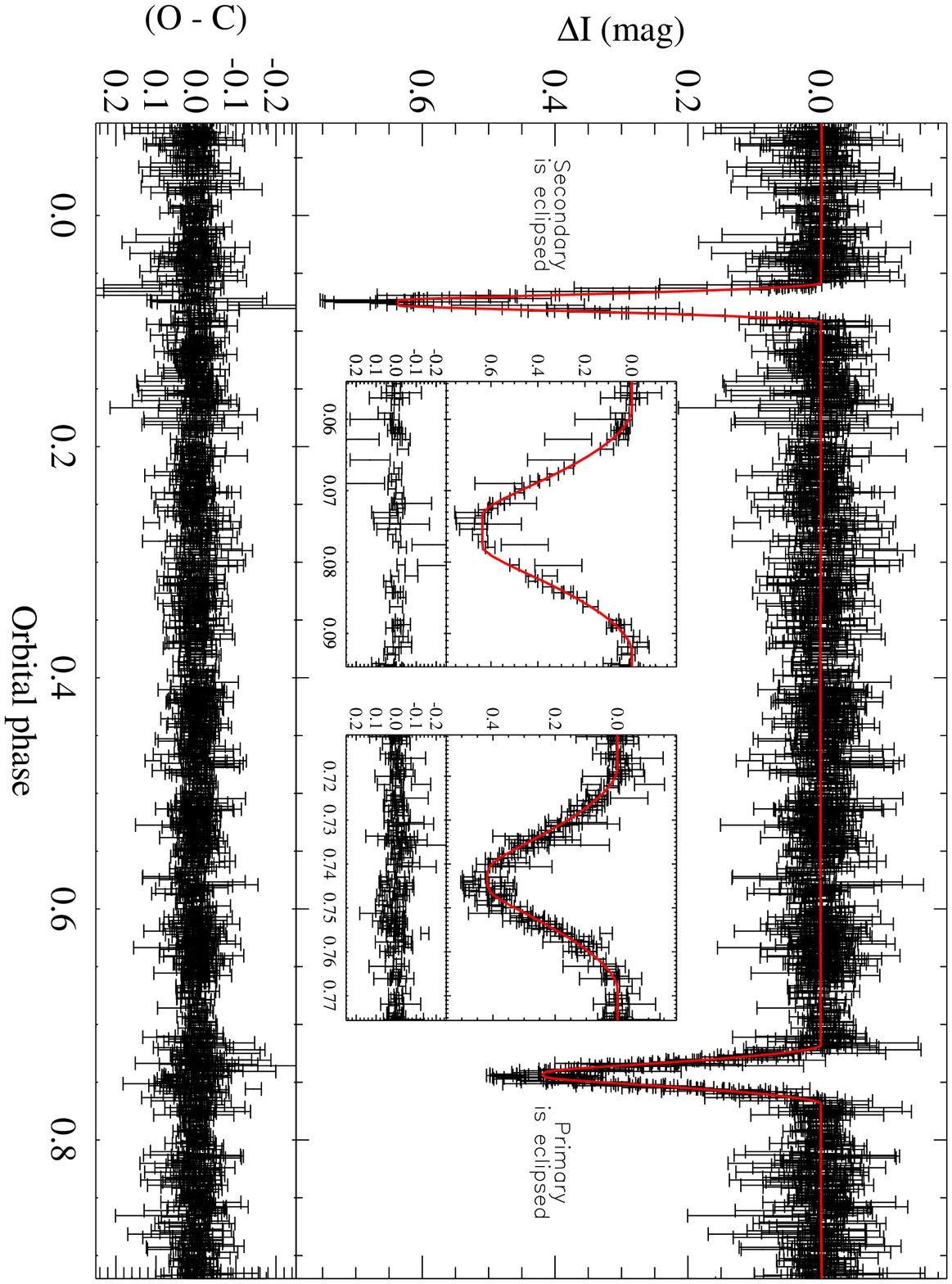]{
\label{Ilc-fig}
$I_C$-band light curve of \bd. The individual photometric
measurements from Table~\ref{lc-data} are plotted, rectified using 
sinusoidal fits to the out-of-eclipse portions of the light curve (see
\S\ref{analysis}), folded on the orbital period and phased to the
time of periastron (see Table~\ref{properties-table}). The solid line
is a \model\ model based on a simultaneous fit to the rectified
$I_C$-band light curve and the {\it Phoenix} radial velocities (see
Table~\ref{rv-data}). Residuals are shown at bottom. Insets show
detail around primary and secondary eclipses, which in this system
correspond to the eclipses of the secondary and primary components,
respectively. The r.m.s.\ residual is 0.02 mag, comparable to the mean
photometric error. The reduced $\chi^2$ of the fit is 1.035. Note:
this figure appears in color in the electronic version of the journal.
}


\figcaption[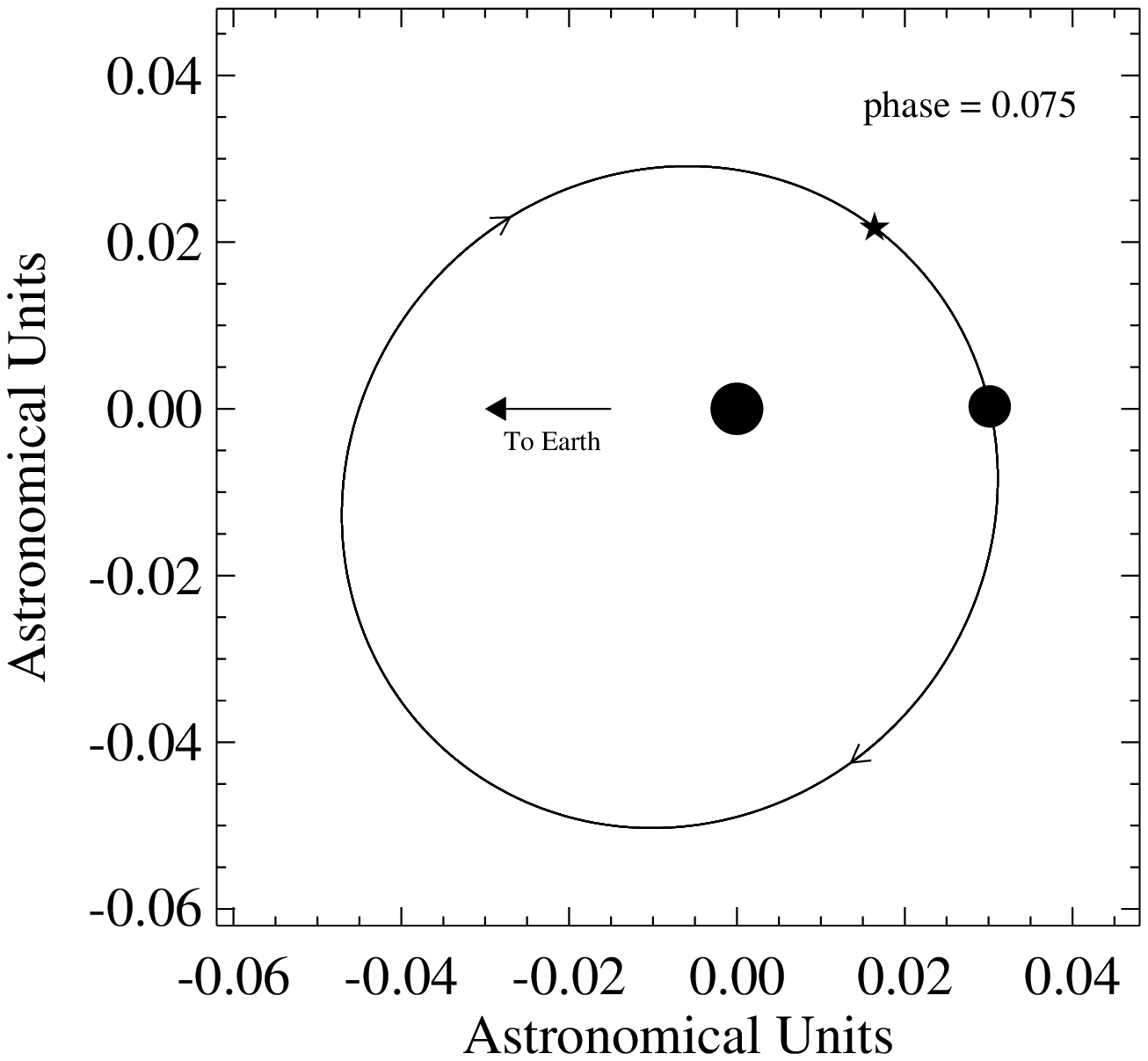]{
\label{orbit-fig}
The geometry of the orbital plane of \bd\ in the rest frame of the
primary (more massive) brown dwarf is illustrated, to scale, at primary
eclipse (orbital phase 0.075; see Fig.\ \ref{Ilc-fig}). The brown
dwarfs are represented by filled circles, their radii also to scale
(the adopted rotation period of $P_{\rm rot} = 3.3$ d constrains the
oblateness of both components to be less than 0.5\%). The position
of periastron (orbital phase 0.0) is indicated by a star symbol, and
small arrows indicate the direction of the secondary's orbit about
the primary. The observer is to the left, as indicated by the arrow.
Note that, at this orbital phase corresponding to the deeper eclipse,
the secondary (less massive) brown dwarf is the one eclipsed.
} 


\figcaption[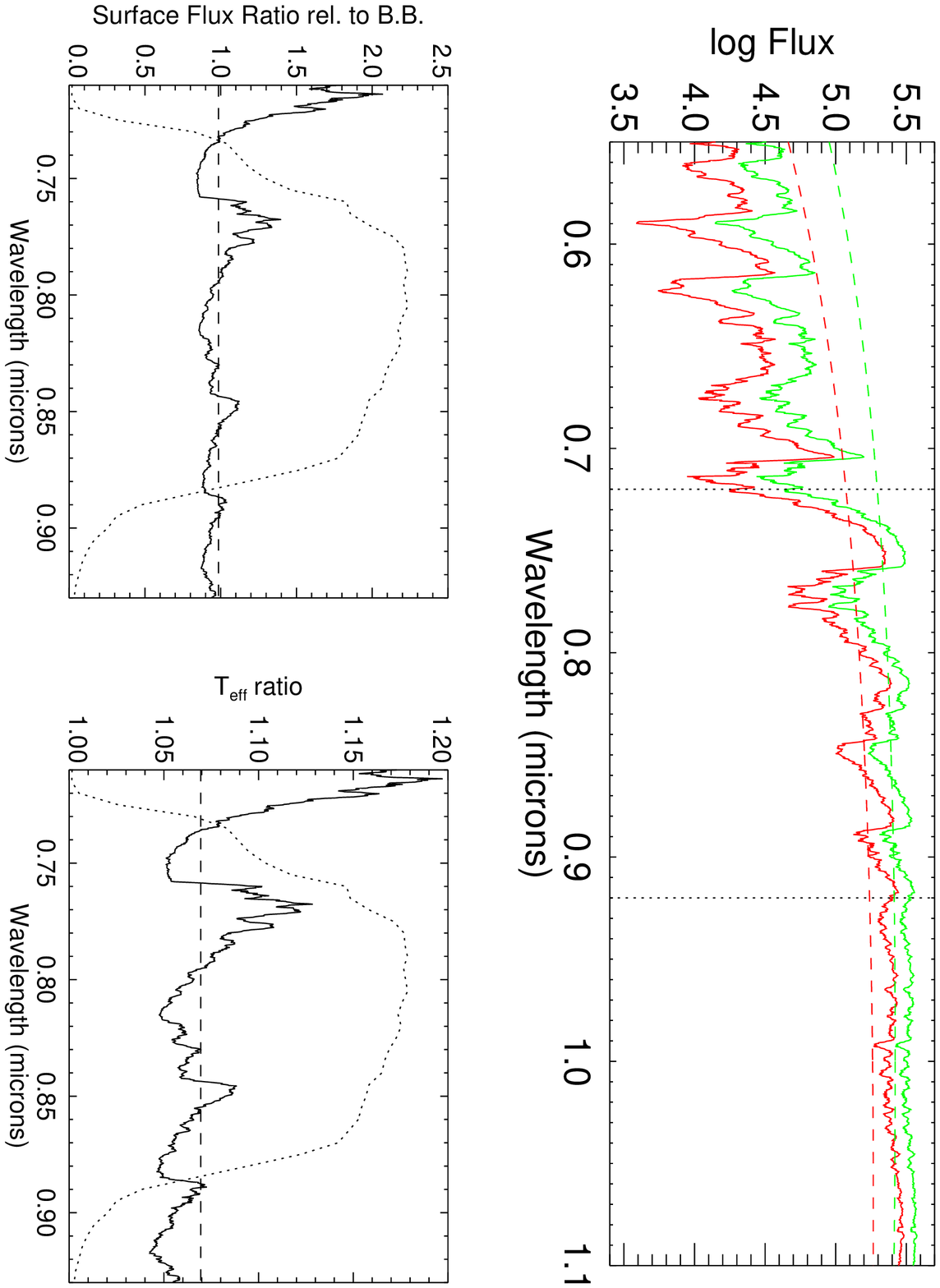]{
\label{speccomp-fig}
We use the {\sc cond} atmosphere models of \citet{allard01} to check
whether the non-blackbody spectra of brown dwarfs may explain the
unexpected reversal of effective temperatures with mass in \bd. {\it
Upper panel}: {\sc cond} model spectra at the effective temperatures
of the primary (green) and secondary (red) components of \bd\ are
compared with simple blackbodies of the same temperatures (dashed
lines). Dotted vertical lines demarcate the approximate bandpass of the
$I_C$ filter used for our light-curve observations. {\it Bottom left}:
The flux ratio of the {\sc cond} models in the upper panel is shown
relative to the blackbody flux ratio as a function of wavelength in
the $I_C$ band. Also
shown is the $I_C$-band filter bandpass profile (dotted curve) and
the bandpass-weighted mean {\sc cond}-to-blackbody flux ratio (dashed
line), which is 1.02 in this case; that is, the {\sc cond} flux ratio
is within 2\% of the blackbody value. {\it Bottom right}: Same as
{\it bottom left}, except showing the $T_{\rm eff}$ ratio implied
by the {\sc cond} flux ratio. The bandpass-weighted value is 1.069
(dashed line), within 0.5\% of the value of $1.064 \pm 0.004$ found in
our \model\ analysis (\S\ref{analysis}) using simple blackbodies. Note:
This figure appears in color in the electronic version of the journal.
} 


\figcaption[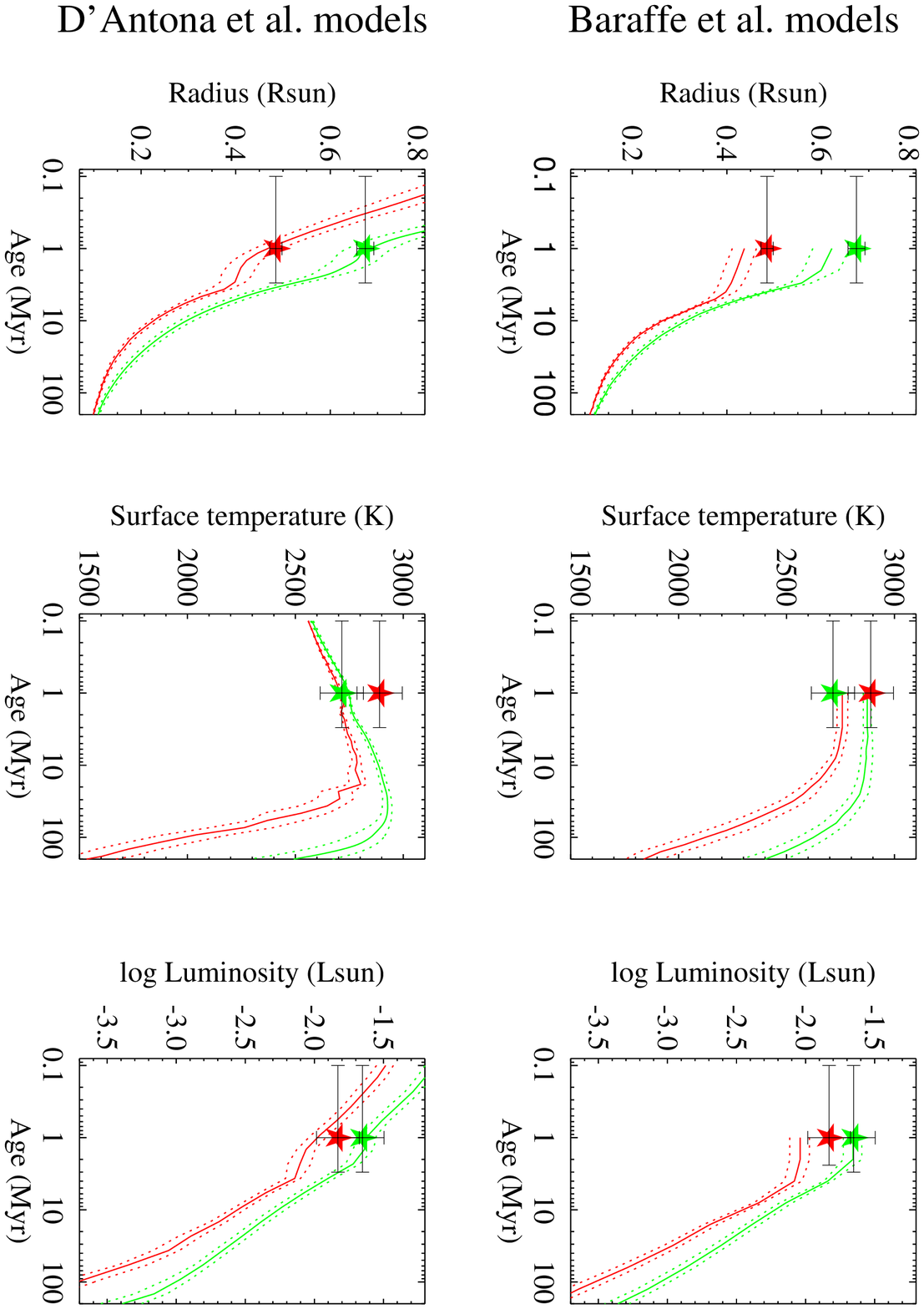]{
\label{models-fig}
Comparisons of observations with theoretical models of young brown
dwarfs. The measured radii, effective temperatures, and luminosities of
\bd\ (symbols with error bars) are compared to the values predicted by
two sets of theoretical models \citep{baraffe98,dantona98} of young
brown dwarfs. Solid curves show the predicted evolution from 0.1 to
100 Myr for brown dwarfs with masses equal to those measured for
the primary (green) and secondary (red) brown dwarfs in \bd. (The
theoretical calculations of \citet{baraffe98} do not extend to
ages less than 1 Myr.) Dashed curves bracketing the solid curves
represent the $1\sigma$ measurement uncertainties in those masses. Observed 
and predicted values are generally in good agreement, particularly with the
\citet{dantona98} models. Note that the effective temperature of the
primary is predicted by both sets of models to be warmer than that of
the secondary at any particular age, but that the \citet{dantona98}
models allow the primary to be cooler than the secondary if it is
sufficiently younger (see also Fig.\ \ref{models2-fig}). Note: This
figure appears in color in the electronic version of the journal.
} 


\figcaption[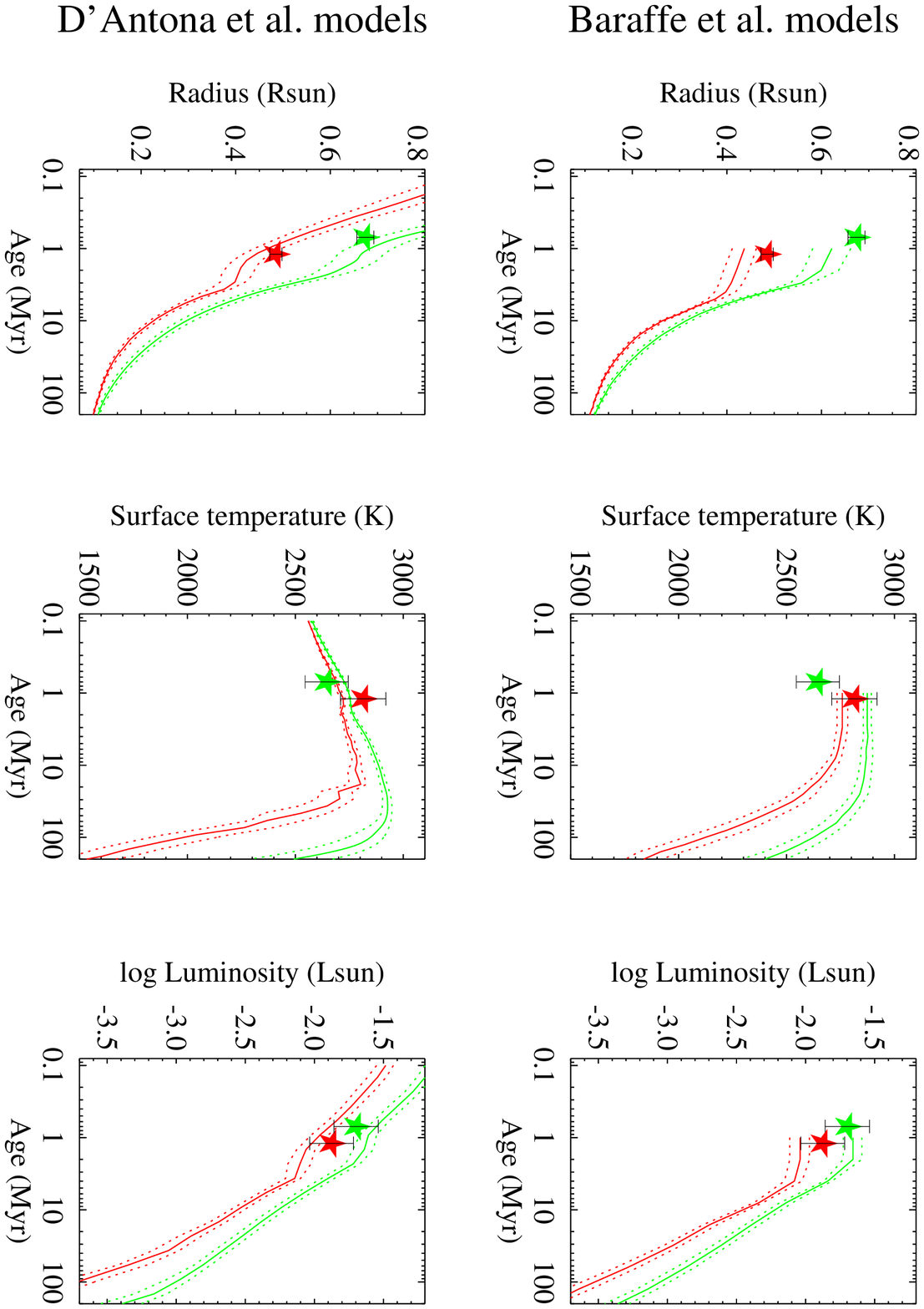]{
\label{models2-fig}
Same as Fig.\ \ref{models-fig}, except with the \bd\ components
arbitrarily separated in age by 0.5 Myr ($\tau_1 = 0.75$ Myr, $\tau_2 =
1.25$ Myr). This demonstrates that the temperature reversal observed in
\bd\ may in fact be marginally consistent with the \citet{dantona98}
models if the \bd\ primary is taken to be slightly younger than
the secondary. Note: This figure appears in color in the electronic
version of the journal.
} 

\clearpage
\begin{figure}
\plotone{f1.ps}
\centerline{f1.ps}
\end{figure}

\clearpage

\begin{figure}
\rotate
\plotone{f2.ps}
\centerline{f2.ps}
\end{figure}
\clearpage

\begin{figure}
\rotate
\plotone{f3.ps}
\centerline{f3.ps}
\end{figure}
\clearpage

\begin{figure}
\plotone{f4.ps}
\centerline{f4.ps}
\end{figure}
\clearpage

\begin{figure}
\plotone{f5.ps}
\centerline{f5.ps}
\end{figure}
\clearpage

\begin{figure}
\rotate
\plotone{f6.ps}
\centerline{f6.ps}
\end{figure}
\clearpage

\begin{figure}
\plotone{f7.ps}
\centerline{f7.ps}
\end{figure}

\end{document}